\documentclass{emulateapj}

\usepackage{graphicx}
\usepackage{color}
\usepackage{ulem}
\usepackage{comment}

\def\HI{{\rm H\,I}}
\def\Mtot{M_{\rm tot}}
\def\Mhi{M_\HI}
\def\hMsun{h^{-1}{\rm M}_\odot}
\def\Msun{{\rm M}_\odot}

\begin{document}
\preprint{FERMILAB-PUB-09-550-A}
\shortauthors{Mar\'in et al.}
\shorttitle{Large Scale $\HI$ Bias}

\title{Modeling the Large Scale Bias of Neutral Hydrogen}
\author{Felipe A. Mar\'in\altaffilmark{1,2}, Nickolay Y.\
 Gnedin\altaffilmark{3,1,2}, Hee-Jong Seo\altaffilmark{3} and Alberto Vallinotto\altaffilmark{3}}
\altaffiltext{1}{Department of Astronomy \& Astrophysics, The
 University of Chicago, Chicago, IL 60637 USA;
 fmarinp@oddjob.uchicago.edu} 
\altaffiltext{2}{Kavli Institute for Cosmological Physics and Enrico
 Fermi Institute, The University of Chicago, Chicago, IL 60637 USA} 
\altaffiltext{3}{Center for Particle Astrophysics,
Fermi National Accelerator Laboratory, P.O. Box 500
Kirk Rd. \& Pine St., Batavia, IL 60510-0500; gnedin@fnal.gov, sheejong@fnal.gov, avalli@fnal.gov}


\journalinfo{Accepted by The Astrophysical Journal}

\begin{abstract}
We present new analytical estimates of the large scale bias of neutral hydrogen ($\HI$)
 We use a simple, non-parametric model which monotonically relates the total mass of a halo $\Mtot$ with 
 its $\HI$ mass $\Mhi$ at zero redshift; for earlier times we assume limiting models for the 
 $\Omega_{\HI}$  evolution consistent with the data presently available, as well as two main scenarios for the evolution
of our $\Mhi$ - $\Mtot$ relation. We find that both the linear and the 
first nonlinear bias terms exhibit a strong evolution
with redshift, regardless of the specific limiting model assumed for the $\HI$
 density over time. These analytical predictions are then
 shown to be consistent with measurements performed on the Millennium Simulation. Additionally,  we
show that this strong bias evolution does not sensibly affect the measurement of the $\HI$ power spectrum.  
\end{abstract}

\maketitle

\section{Introduction}

The 21cm emission line of neutral hydrogen has been the workhorse of
astronomy for over 50 years. Its use for galactic astronomy cannot be
overestimated, and it now promises to revolutionize cosmology as well
\citep[cf.][and references therein]{21cm:fob06,21cm:bl07,21cm:pl08,Mao:08,Visbal:08}.

Compared to present probes of large-scale structure, the 21 cm
intensity mapping technique \citep{21cm:cppm08,21cm:wl08} offers
three advantages. First, 
it allows to reconstruct the spatial distribution of $\HI$ in the
universe over volumes much larger than the ones currently probed by
galaxy redshift surveys. Since the precision with which the power
spectrum of density fluctuations can be measured is approximately
proportional to the number of independent Fourier modes that fit into
the survey volume, a larger survey volume leads to higher precision
in the measured power spectrum and hence cosmological parameters 
\citep{Loeb:2008hg}. Second, 21 cm intensity
mapping allows to probe a very wide range of redshifts and to
map $\HI$ distribution in space and time not only for the epoch
after reionization ($z\lesssim 6$) but also during reionization and
even in the so-called dark ages, before the first galaxies were
formed ($z\gtrsim 20$). Third, using radio telescopes is one of the most
economical ways to probe the dark energy evolution among all
current and planned experiments, although several crucial
advances in calibration and foreground removal must be
made for the intensity mapping approach to reach its stated goals.  

As a dark energy probe, 21 cm intensity mapping is primarily useful
for measuring the baryon acoustic oscillations (BAO) in the matter power
spectrum \citep{Wyithe:2007rq,Chang_et_al:08,21cm:wl08,Ansari:08,Seo_et_al:09},
which provides important information on constraining cosmological
models and parameters. Measurements of the scale of the BAO in the matter power spectrum as a function of redshift, 
when combined with the true physical scale, i.e., the sound horizon scale measured from the 
cosmic microwave background (CMB), probe the angular diameter distances and Hubble parameters to various redshifts,
 and therefore dark energy properties \citep[see, for instance, ][]{hu_white:96, eisenstein_etal:98, eisenstein:02, blake_glazebrook:03, linder:03,hu_haiman:03,
 seo_eisenstein:03}. The distinct oscillatory feature of the BAO allows us to isolate
 and measure the BAO information independent of the overall broadband shape of the HI power spectrum. 
Therefore in principle BAO measurements do not require a knowledge of the HI bias, i.e., the relation 
between the neutral hydrogen and the total matter clustering, which affects the overall shape of the HI power spectrum. 

However, understanding the HI bias is important for the 21cm BAO surveys in the following ways. 
First the prediction of the signal to noise of the BAO measurements is sensitive to the knowledge of the 
clustering bias of HI \citep{Chang_et_al:08,Seo_et_al:09}. Second, understanding HI bias will help us to extract cosmological
 information from the broadband shape of the power spectrum in addition to the BAO. Third, the nonlinear effects on the BAO, such as the degradation and the shift of the feature, may differ for different biased tracers \citep[e.g.,][]{padmanabhan_white:09, metha_etal:10}. Knowing the properties 
of HI bias will allow us to approximately estimate the degree of the expected nonlinear effect on the BAO in the HI distribution.

In this work, we focus on two
critical aspects of the large-scale $\HI$ bias: its dependence on the relation between
 $\HI$ mass and halo mass and its evolution with redshift.
To study the first aspect,  we use a non-parametric model where
we assume a one-to-one correspondence between a  $\HI$ mass 
present in a dark matter halo $\Mhi$ 
and the total mass $\Mtot$ of this halo. 
Using this $\Mhi -\Mtot$ relation we
estimate the large-scale bias of the $\HI$ using  a
model containing elements on the halo occupation distribution (HOD) formalism 
and the $\HI$
mass function measured at $z=0$ \citep{igm:zmsw05}.
We then use this relation to paint $\HI$ on
Millennium Simulation halos \citep{springel_etal:05} and to measure the
power spectrum of neutral hydrogen in order to test our model assumptions.

With respect to the redshift evolution of the bias, it is absolutely critical
to know how  the $\HI$ mass function evolves with time.
However, to the best of our
knowledge, little is known about the redshift evolution of $\HI$ 
both on the theoretical and the observational sides \citep[but see][]{wyithe_etal:09}. A
detailed study of the bias evolution requires two extra pieces of
information which are currently poorly constrained by observational
data: the evolution of the total HI mass in the universe and the
evolution of the HI mass function. For both of these, we empirically
assume limiting evolution models consistent with the few available
data that should bracket the actual evolution. We then proceed to
predict the bias evolution in these scenarios, being aware of the fact that
the actual evolution will lie somewhere in between these limiting
cases.

The paper is organized as follows. In \S\ref{app:biashod} we introduce the basics of our model, 
we derive the relation $M_{\rm HI}-Mtot$ at $z=0$, we introduce the limiting cases for its evolution
 and we define the bias. In \S\ref{section:results} we present our main results: analytic predictions
 for the evolution of HI bias, measurement of its scale dependence on the Millennium Simulation and how it 
would affects measurements of the power spectrum. Finally, we conclude in \S\ref{section:conclusions}. 
To compare analytical and numerical results, throughout the paper we use  a flat $\Lambda$CDM cosmology with 
$\Omega_M=0.25$,  $\sigma_8=0.9$,
Hubble parameter $h=0.73$ and initial power spectrum index of $n_s=1$, consistent with the cosmological 
parameters used in the Millennium
Simulation.

\section{Background}
\label{app:biashod}

\subsection{Basic  formalism}

The HOD model \citep[cf.][and references therein]{cooray_sheth:02,berlind_weinberg:02} 
provides a
parameterized prescription for the spatial distribution of objects that
can be found inside dark matter halos. Originally inspired to model
the distribution of galaxies, it can also be used to model the spatial
distributions of other astronomical objects, such as the neutral
hydrogen clouds. The advantage in using this approach is that the
problem of deriving the spatial distribution naturally falls into two separate parts: the first part is
concerned with the distribution of halos in the universe, and the
second part involves how objects (i.e., galaxies, hydrogen clouds,
etc) populate these halos. In this work we use some of the elements
of this HOD formalism to attempt to estimate the large-scale bias, i.e. 
in a quasi-linear regime, where the estimations of the 21 cm correlations
for studying the BAO will take place.

The first component is a model for the 
distribution of dark halos in the universe.
In the analytic approach, one specifies the halo mass function, $n(M)$, 
the spatial correlations of the halos, and the density profiles for 
halos of mass $M$, all based on halo collapse models and fits to $N$-body simulations.

To describe the mass function of dark matter halos (the spatial density of halos as a 
function of mass), we use the Sheth \& Tormen mass function \citep{sheth_et_al:01, sheth_tormen:02}:
\begin{equation}
n_h(M) = \frac{\bar{\rho}_m}{M^2}\left|\frac{d\ln\sigma}{d\ln M}\right| f(\nu),
\end{equation}
where $\bar{\rho}_m \equiv \Omega_m \rho_c$ is the average matter density in the universe at the
epoch of observation,  $\nu(M)$, defined by 
$\nu(M)\equiv\delta_c/\sigma(M,z)$, is the ratio between the threshold for growth
of linear fluctuations $\delta_c=1.69$ and the rms variance of a sphere of radius 
$R$  (corresponding to a mass $M$) at redshift $z$
\begin{equation}
\sigma^2(R,z)= \int \frac{d^3k}{(2\pi)^3} \left| W(k,R) \right|^2P(k,z).
\end{equation}
Here, $P(k,z)$ is the linear matter power spectrum and $W(k,R)$ is the 
Fourier transform of a top-hat window of radius $R$. The function $f(\nu)$ motivated by the
ellipsoidal collapse model  \citep{sheth_tormen:02} is
\begin{equation}
f(\nu) = A\sqrt{\frac{2a\nu^2}{\pi}}\left[1+(a\nu^2)^{-p}\right]e^{-a\nu^2/2},
\end{equation}
where for normalization purposes, $A=0.322$, $p=0.3$ and $a=0.707$. 

The second component is a model specifying how the
objects under consideration (in the present case, the amount of neutral hydrogen)
occupy the dark halos as a function of halo mass. In a HOD model,
for the largest scales, 
we would be  interested in the average number of galaxies $\left< N(\Mtot)\right>$ 
that populate a particular halo of mass $\Mtot$. In our case, we are interested in
the average number of $\HI$ atoms, or equivalently the average mass
of $\HI$ inside a halo. For the scope of this paper, which deals with large scale-bias, we
are not using the complete HOD prescription: this would mean specifying the 
complete probability distribution function of $P(M_{\HI}|\Mtot)$, where its higher
moments contribute to the estimation of the bias on small, nonlinear scales; for large scale
bias, we only need the average $\HI$ mass $\left<M_\HI(\Mtot)\right>$. When interested in the
small-scale bias,  details the spatial and velocity distribution of the $\HI$ inside the halos are needed; these
are out of the scope of this paper (see for instance \citealt{bagla_khandai:09}, \citealt{obreschkow_etal:09} \citealt{wyithe_etal:09}
for treatments that include small-scale clustering).

In the following subsection, we explore different models for the neutral hydrogen mass function
and the relations between the neutral hydrogen $\Mhi$ and the total mass of a halo $\Mtot$.

\subsection{Neutral Hydrogen Mass Function}
\label{section:h21mf}

The neutral hydrogen mass function of galaxies in the local universe
has been measured by the HIPASS survey \citep{igm:zmsw05}. However,
the masses of the halos that host galaxies with a given  $\HI$ mass are not,
generally, known. Thus, we need to construct a plausible model for the
relation between the total mass of a halo $\Mtot$ and its
neutral hydrogen mass $\Mhi$.

While such a relation does not have to be a simple function, models
that assume a one-to-one correspondence between the galaxy luminosity
and the halo mass
\citep{colin_etal:99,kravtsov_klypin:99,kravtsov_etal:04} provide
remarkably good fit not only to the galaxy correlation functions
\citep{qlf:cwk06,marin_etal:08} but also to a wide variety of
observational tests
\citep{nagai_kravtsov:05,ng:g08,ng:vg09,misc:gwlb09,conroy_wecshler:09,tinker_conroy:09}.
Therefore, it is tempting to assume such a relation between the $\HI$
mass and total halo mass as well, that is, to assign one HIPASS source
to one dark matter halo.

There is one complication with such an approach -- not all galaxies contain
$\HI$. Thus, we need to account for the fraction of galaxies that
do. Unfortunately, there are no reliable observational measurements of
the fraction of $\HI$-rich galaxies as a function of galaxy mass or
luminosity. Instead, we adopt a simple ansatz that all blue galaxies
contain $\HI$ and all red galaxies do not. Then, we can use the
abundance matching technique if we match the abundances of HIPASS and
blue galaxies.

Specifically, the relation between the total mass of a
dark matter halo $\Mtot$ and its $\HI$ mass $\Mhi$ can be obtained by
matching the two cumulative mass functions,
\begin{equation}
 n_\HI(>\Mhi) = n_B(>\Mtot),
 \label{eq:nn}
\end{equation}
where $n_\HI$ is the observed cumulative mass function
\citep{igm:zmsw05} and
\begin{equation}
 n_B(>\Mtot) = \int_{\Mtot}^{\infty} dM f_B(M) n_h(M).
\end{equation}
In the last equation, $f_B$ is the fraction of blue galaxies as a
function of halo mass.

\begin{figure}[t]
\plotone{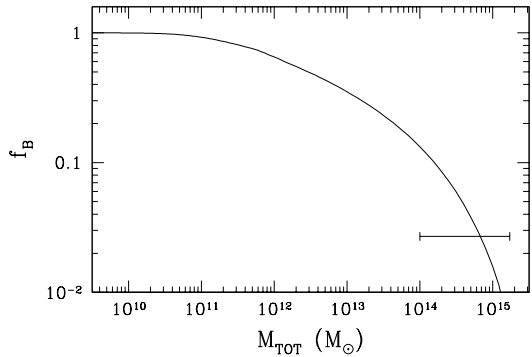}
\caption{\label{fig:fblue}
Fraction of blue galaxies as a function of total halo mass from
matching the cumulative halo mass function and the observed SDSS
galaxy luminosity function. The horizontal range shows the observed
fraction on blue BCGs in the maxBCG galaxy cluster catalog (E. Rykoff,
 private communication).}
\end{figure}

That fraction can be computed from the measured Sloan Digital Sky Survey (SDSS)
red and blue galaxy luminosity functions \citep{Montero_prada:09}
using the same abundance-matching technique. Namely, using the total
SDSS luminosity function from \citet{Montero_prada:09} we first match
halo masses to $r$-band magnitudes as
\begin{equation}
  n_{\rm gal}(<m_r) = \int_{\Mtot}^{\infty} dM n_h(M);
\end{equation}
such a matching gives us a relation between a galaxy magnitude $m_r$ and
the halo mass $\Mtot$. Then, for each $m_r$ (and, hence, $\Mtot$) we
compute the fraction of blue galaxies from the measured blue galaxy
luminosity function \citep{Montero_prada:09}. The so-measured blue
galaxy fraction is shown in Figure \ref{fig:fblue} together with am
estimate of the blue BCG fraction from the maxBCG galaxy cluster
catalog (E. Rykoff, private communication). Estimates of the blue BCG
fraction in X-ray selected clusters usually give substantially larger
numbers \citep{Bildfell_etal:08}; however, because X-ray emission is
likely to correlate with the existence of a cooling flow (and,
hence, a blue BCG), the latter estimates should be taken as upper
limits (E. Rykoff, private communication).

\begin{figure}[t]
\plotone{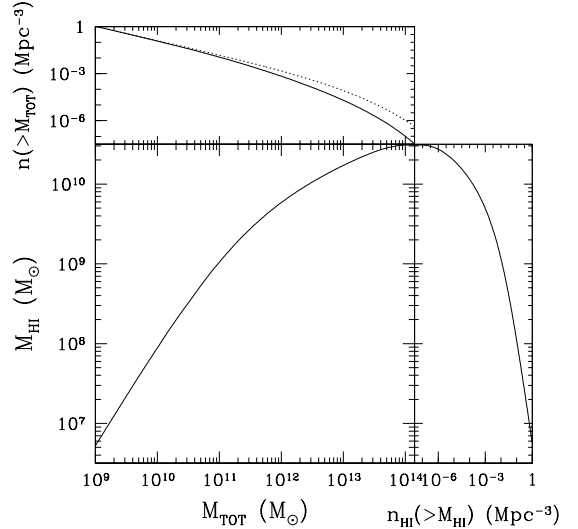}
\caption{\label{fig:himf}
Cumulative mass function of dark matter halos of blue galaxies (top panel), the
cumulative $\HI$ mass function of galaxies \citep[from][ right
 panel]{igm:zmsw05}, and the relation between the galaxy $\HI$
mass and its halo mass (central panel) obtained by matching the two,
at $z=0$. The dotted line in the top panel shows the cumulative mass
function of all dark matter halos, hosting both blue and red galaxies.}
\end{figure}

The resulting matching between the $\HI$ mass of a galaxy and its
total mass from Equation (\ref{eq:nn}) is shown in Figure
\ref{fig:himf}. The relation between $\Mhi$ and $\Mtot$
 from the central panel allows us to compute the clustering
properties of $\HI$ from the known clustering properties of the dark
matter halos.

\begin{figure}[t]
\plotone{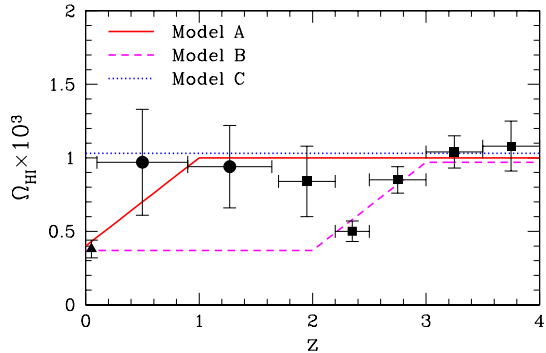}
\caption{\label{fig:models}
Three models (A, B, and C) for the redshift evolution of the total
$\HI$ density that we explore in this paper. The lines for the models
are slightly shifted vertically for clarity -- all models adopt
$\Omega_\HI=10^{-3}$ at $z>3$. Black symbols with error
bars show the observational measurements from \citet{igm:zmsw05}
(triangle), \citet{igm:rtn06} (circles), and \citet{igm:phw05}
(squares).}
\end{figure}

Unfortunately, little is known about the $\HI$ mass function beyond
$z\approx0.1$. To obtain the $\Mhi$ - $\Mtot$ matching at higher
redshifts, we therefore must extrapolate the observed $\HI$ mass
function to earlier times. This extrapolation is, of course, not
unique and it requires two ingredients: the evolution of the total
$\HI$ mass and how the $\HI$ mass is distributed among halos. 
To make
the matter even worse, the observational constraints on the total
$\HI$ mass density in the universe $\rho_{\HI}$ (or, equivalently, the
cosmological density parameter $\Omega_\HI$) are imprecise. Figure
\ref{fig:models} shows a representative sample of these constraints
for a range of redshifts \citep{igm:zmsw05,igm:phw05,igm:rtn06}. While
at $z\approx0$ and $z\ga2$ the measurements are reasonably robust, at
intermediate redshifts the error bars are large and the behavior is
non-monotonic. Therefore, to explore a wide enough range of possible
evolutionary histories of $\Omega_{\HI}$, we choose three
representative models.

\begin{itemize}
\item Model A provides a reasonable fit to all data points except for
a single $z\approx2.3$ point from \citet{igm:phw05}.
\item Model B follows the suggestion of \citet{igm:pw09}, who argue
that $\Omega_\HI$ does not evolve between $z=0$ and $z\approx2$.
Since it provides a plausible lower limit to the redshift evolution of
$\Omega_\HI$, we use this model even if it ignores the measurements of
\citet{igm:rtn06}.
\item In the same spirit, Model C serves as an upper limit to the
possible $\Omega_\HI(z)$. Formally, it is inconsistent with the HIPASS
measurement, but we assume that $\Omega_\HI$ in Model C increases
rapidly from the HIPASS value of $3.8\times10^{-4}$ to about $10^{-3}$
within a redshift interval $\Delta z \ll 1$.
\end{itemize}

To fully specify the $\HI$ mass function we then need to address how
the total $\HI$ mass is distributed among halos as a function of
redshift. If the shape of the $\HI$ mass function is approximately
preserved at $z>0$, then 
\begin{equation}
  n_\HI(>\Mhi,z) = q_N(z) n_0(>q_M(z) \Mhi),
\end{equation}
where $n_0(>\Mhi)$ is the observed cumulative HIPASS mass function
\citep{igm:zmsw05} at $z=0$. The two quantities $q_N$ and $q_M$ are not
independent; for a given evolutionary history of neutral hydrogen
contents of the universe $\rho_\HI(z) \equiv \Omega_\HI(z) \rho_{\rm
  crit}$, $q_N$ and $q_M$ are related by the constraint 
\begin{eqnarray}
  \rho_\HI(z) & = & \int_0^\infty M_\HI
  \frac{dn_\HI(\Mhi,z)}{dM_\HI}\, dM_\HI \nonumber \\
  & = & \int_0^\infty n_\HI(>\Mhi,z)\, dM_\HI \nonumber \\
  & = & \frac{q_N(z)}{q_M(z)} \rho_\HI(0).
\end{eqnarray}
Obviously, $q_N(0) = q_M(0) = 1$.

It is not practical to explore all possible evolutionary histories
for $n_\HI(>\Mhi,z)$; therefore, we restrict this work to only two scenarios.	

\begin{description}
\item[Pure Number Density Evolution] scenario (hereafter, PNE) assumes
  that the $\HI$ masses of individual halos do not change at high
  redshift ($q_M=1$ for all $z$), but only the number density of halos
  of a given mass changes to reflect the change in the total neutral
  hydrogen content of the universe, $q_N =
  \Omega_\HI(z)/\Omega_\HI(0)$.
\item[Pure Mass Evolution] scenario (hereafter, PME --a direct analog
  of a ``pure luminosity evolution'' scenario in optical surveys)
  adopts an opposite extreme where the number density of halos remain
  the same ($q_N=1$ for all $z$), but, instead, their masses change,
  $q_M = \Omega_\HI(0)/\Omega_\HI(z)$.
\end{description}

The justification for such a restriction is as follows. If the number
density of galaxies with a given $\HI$ mass increases monotically with
increasing redshift, then $q_N(z)$ is a monotonic function of $z$ (which is
expected, since galaxies merge but do not break up). If, for a
given halo, its $\HI$ also increases monotically with $z$ (which is
expected, if $\HI$ gets consumed by star formation and new accretion
never exceeds the consumption), then $q_M(z)$ is also a monotonic
function of $z$. Since $q_N(0) = q_M(0) = 1$, our two scenarios
bracket all possible evolutionary path for $q_N$ and $q_M$.

Obviously, these assumptions do not hold all the time --for example,
for some galaxies accretion of fresh gas can exceed the consumption by
star formation at some times. However, over long timescales and over
the whole galaxy population, it appears plausible that our two
scenarios bracket the majority of evolutionary histories for
$n_\HI(>\Mhi,z)$.

One more ingredient of our model can evolve with redshift --and that
is the fraction of blue galaxies $f_B(\Mtot)$. The measured fraction
of blue galaxies at $z=0$ from Figure  \ref{fig:fblue} is exceptionally
well (within the thickness of the line in Figure \ref{fig:fblue})
fitted by the following simple formula:
\begin{equation}
  f_B(\Mtot) =\left(1+\frac{\Mtot}{M_1}\right)^{-0.25}\exp\left(-\left[\frac{\Mtot}{M_2}\right]^{0.6}\right)
  \label{eq:fblue}
\end{equation}
with $M_1=2.5\times10^{11}\Msun$ and $M_2=3.0\times10^{14}\Msun$. It
is tempting to associate $M_1$ with the characteristic mass of dwarf
ellipticals, and $M_2$ with the characteristic mass of a galaxy
cluster. Then the behavior of $f_B$ with $M$ becomes well justified
physically: for $\Mtot \gtrsim M_1$ a small fraction of all halos host
dwarf and (at larger masses) regular elliptical galaxies, and $f_B$
becomes a gradually decreasing function of mass. But for $\Mtot
\gtrsim M_2$ halos turn into galaxy groups and clusters, and the
fraction of halos dominated by blue galaxies plummets.

With this plausible (although by no means proven) interpretation, we
can come up with reasonable evolutionary scenarios for $f_B$ as a
function of $z$. Specifically, we adopt the parametric form
(Equation (\ref{eq:fblue})), but allow $M_1$ and $M_2$ to evolve with redshift as
\begin{eqnarray}
  M_1(z) & = & 2.5\times10^{11}(1+z)^\alpha \Msun, \nonumber \\
  M_2(z) & = & 3.0\times10^{14}(1+z)^\beta \Msun.
  \label{eq:fblueofz}
\end{eqnarray}
It seems plausible to assume that the fraction of blue galaxies at a given halo
mass is higher at high redshift, i.e., $\alpha\geq0$ and $\beta\geq0$. To investigate the dependence of our results on these two parameters, we consider below three representative cases with $(\alpha,\beta)$ = $(0,0)$, $(0,1)$, and $(1,1)$. This set of values is, of course, not exclusive, but it is sufficient to evaluate the robustness of our results to these parameters.

In this work we therefore consider a range of possibilities --our models
A, B, and C for $\Omega_\HI(z)$ and two evolutionary scenarios (PNE
and PME) --as sampling a plausible range of the redshift dependence of the $\HI$ mass function. It seems reasonable to speculate that the actual evolution should lie somewhere in between these different scenarios.

\begin{figure*}
\plotone{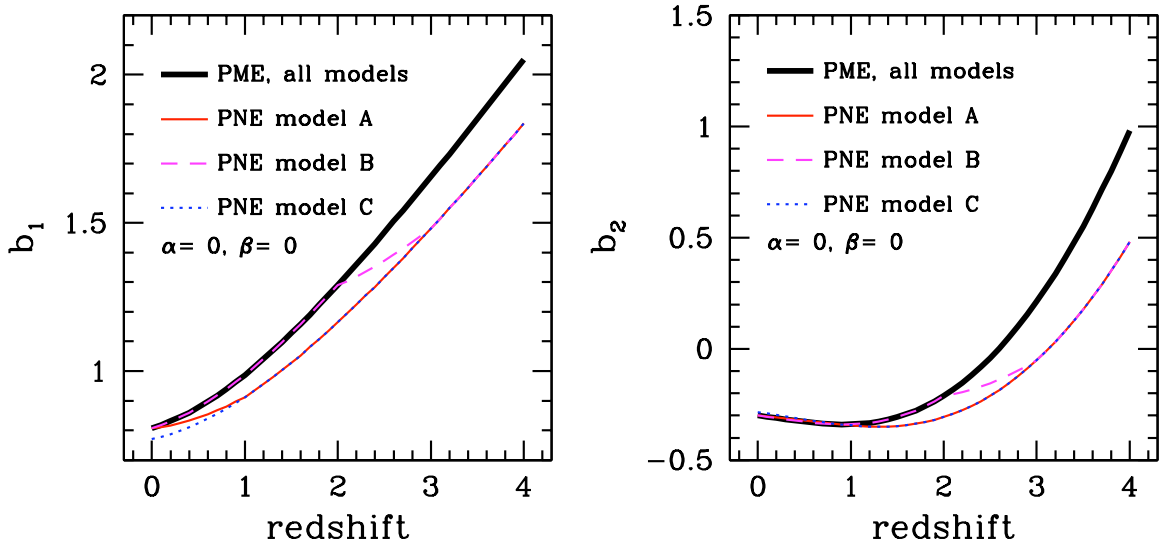}
\caption{Bias coefficients of neutral hydrogen clouds as a function of redshift. 
\textit{Left:}  linear bias $b_1$ vs. redshift in the pure mass evolution scenario 
for all models (thick black solid line), and the pure number evolution scenario 
for model A (thin red solid line),  model B (dashed magenta line) and model
C (blue dotted line).  \textit{Right:} first nonlinear bias term $b_2$ vs. redshift 
for the same models described in the left plot. All coefficients are calculated
with $M_{\rm min}=10^4$ $\hMsun$.}
\label{fig:biasz}
\end{figure*}

\subsection{Large-Scale Bias of $\HI$}
\label{sec:bias_in_HOD}

On large scales, and not considering
redshift distortions, the $\HI$ bias parameters provide the relations
between the $\HI$ and dark matter correlation functions  \citep{pan_szapudi:05}
\begin{eqnarray}\label{eq:bias-pcf}
\xi_{HI} &\approx& b_{1,\HI}^2\xi_{\rm dm},\\
\zeta_{HI} &\approx& b_{1,\HI}^3\zeta_{\rm dm} + b_{2,\HI}b_{1,\HI}^2(
\xi_{\rm dm}^{(1)}\xi_{\rm dm}^{(2)} + \textrm{perm.}),
\end{eqnarray}
where $\xi$ and $\zeta$  represent the 2-point and 3-point correlation functions, respectively. 
Similar relations are found for quantities calculated in Fourier space (power spectrum and
bispectrum).

To calculate the large-scale bias of neutral hydrogen, we start by considering first
the bias of dark matter halos with respect to the total matter
clustering. In the 
Sheth--Tormen formalism, the bias parameters are prescriptions
obtained from the spherical collapse model. For halos of mass $M$, the
linear and (first) nonlinear 
terms are given by \citep{scoccimarro_etal:01}
\begin{eqnarray}
b_1^h(M) &=& 1+\epsilon_1+E_1,\\
b_2^h(M) &=& \frac{8}{21}\left(\epsilon_1+E_1\right)+\epsilon_2+E_2,
\end{eqnarray}
with coefficients given by
\begin{eqnarray}
\epsilon_1 &=& \frac{a\nu^2-1}{\delta_c},\\
\epsilon_2 &=& \frac{a\nu^2(a\nu^2-3)}{\delta_c^2},\\
E_1 &=& \frac{2p}{\delta_c(1+(a\nu^2)^p)},\\
\frac{E_2}{E_1} &=& \frac{1+2p}{\delta_c}+2\epsilon_1,
\end{eqnarray}
where $i=1,2$ represent the linear and first nonlinear terms, where
their redshift dependence is  encompassed in $\nu=\delta_c/\sigma(M,z)$, which depends
on the redshift evolution of the linear power spectrum.

Using the $\Mhi-\Mtot$ relation obtained in the previous section, 
the $\HI$ bias from all halos with a mass greater than $M_{\rm min}$ is given by the following integral:
\begin{equation}
b_{i,\HI} = \frac{1}{\rho_{\HI}}\int_{M_{\rm min}}^{\infty} dM
n_h(M)b_i^h(M)\left<M_{\HI}(M)\right>.
\label{eq:bhi}
\end{equation}
where $\rho_{\HI}$ is the density of neutral hydrogen:
\begin{equation}
\rho_{\HI}= \int_{M_{\rm min}}^{\infty} dM n_h(M)\left<M_{\HI}(M)\right>.
\label{eq:rhohi}
\end{equation}

In what follows we present the analytic results for the large-scale bias and its evolution in
redshift.

\section{Results}
\label{section:results}

\subsection{Analytic Results}
\label{section:analytic}

We calculate
the bias parameters as a function of redshift for all models described in \S\ref{section:h21mf}. 
In Figure \ref{fig:biasz} we show the results for the models where the galaxy blue fraction does not
evolve with redshift, i.e., in Equation (\ref{eq:fblueofz}), $\alpha=\beta=0$.
In the left panel we present the linear ($b_1$) bias  redshift evolution; 
in the right panel, the nonlinear ($b_2$) bias evolution, with $M_{\rm min}=10^{4}$ $\hMsun$,
which, given the fact that it is not expected that low mass halos (below $10^{7}$ $\hMsun$) contain
significant amounts of $\HI$, is set to the value mentioned above as a conservative limit.
 The black thick line shows the evolution of bias
in the PME scenario; as for the PNE scenario,
the red thin solid line shows the results for model
A, the magenta dashed line shows results for model B, and the blue dotted line does it for model C. 

The first feature that arises from these analytical results is the strong
dependence of the bias terms on the redshift for all the models considered.
Given that $\Omega_{\HI}$ does not evolve significantly  with redshift, the reason for this behavior
lies in the fact that preserving the matching relations (Equation (\ref{eq:nn})) at higher
redshift has the  consequence that 
the most massive halos, which are less numerous at high redshift (i.e., are more biased
with respect to the dark matter distribution), carry more $\HI$ mass
than the halos with the same $\Mtot$ at $z=0$. Therefore, it is expected for both 
$b_1$ and $b_2$ to show a strong redshift dependence.

For the PNE scenario, we can see differences in both bias parameters,
though they are relatively 
small. Between models A and B, the differences are increasing with
redshift, with a peak around $z\sim 2$ --at the moment where
$\Omega_\HI(z)$ of model B is the lowest compared to the rest of the
models; since the fraction of neutral hydrogen in that model is low,
in the PNE scenario this implies fewer halos, and therefore the bias
is expected to be larger. In model C, the difference in the bias evolution compared
to model A appears only at low
redshifts, since it starts with a higher $\Omega_\HI$ and therefore having a smaller bias
on redshifts close to $z=0$. The differences
for the nonlinear bias term $b_2$ are qualitatively similar but much milder.

As opposed to the PNE scenario, there are no differences between the models biases
 in the PME scenario; this
is not surprising since, by definition, in that scenario just the
masses of halos change, but not their total number density. Hence,
irrespective  of the $\Omega_\HI(z)$ evolution, the behavior of the
bias here depends only on the cosmological evolution.
When we compare both scenarios, the differences in the bias parameter
$b_1$, albeit small in general, grow steadily with redshift reaching 10\% level
between $z=1$ and $4$; note here that there are differences between 
the PME scenario and the model B in the PNE scenario only after 
$z \sim 2$. 
For $b_2$ between $z=0$ and $1$ there are almost no
differences between the models, but after that the differences grow
until reaching almost a factor of 2 by $z=4$. In both cases the
PME scenario has a larger bias; this difference is due to the fact that 
more halos are needed in the PNE scenario to account for
the change of neutral hydrogen in the universe, which leads to a lower
halo mass for a given $\HI$ mass and hence a smaller bias than in
the PME scenario. 

\begin{figure*}
\plotone{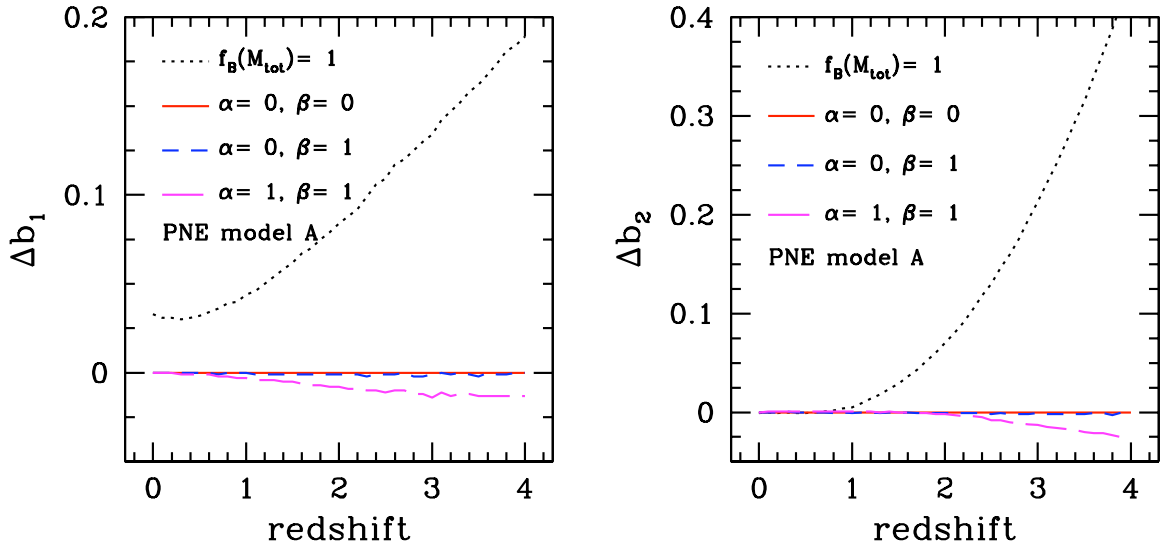}
\label{fig:deltab}
\caption{Differences in the bias parameters (linear $b_1$ in left panel, nonlinear $b_2$in right panel) 
between different blue fraction redshift evolution models. The FID
model is the model A in the PNE scenario, with $(\alpha,\beta)=(0,0)$, in red straight line. \textit{Left:} 
$\Delta b_1 = b_1-b_{1,\textrm{FID}}$, where the black dotted line represents the $f_B(M,z)=1$, the blue 
short-dashed line represents the evolution model with $(\alpha,\beta)=(0,1)$ , and
the long-dashed line shows the results for the blue fraction evolution model with $(\alpha,\beta)=(1,1)$.
\textit{Right:} $\Delta b_2 = b_2-b_{2,\textrm{FID}}$, lines represent the same models of the left panel.}
\end{figure*}

As for the differences between the different blue fraction evolution models, and
the case where we simply assume all galaxies contain $\HI$, we show in 
Figure \ref{fig:deltab} the quantity
\begin{equation}
\Delta b_i = b_i-b_{i,\textrm{FID}},
\label{eq:deltab}
\end{equation}
where $i=1,2$, and the fiducial (FID) model corresponds to the 
model A in the PNE scenario, with $(\alpha,\beta)=(0,0)$. 
The black dotted line represents the $f_B(M,z)=1$, i.e. all galaxies carry $\HI$;
 the blue 
short-dashed line represents the evolution model with $(\alpha,\beta)=(0,1)$ , and
the long-dashed line shows the results for the blue fraction evolution model with $(\alpha,\beta)=(1,1)$,
both in the model A of the PNE scenario. For other models and scenarios the results are qualitatively 
similar: between the different blue fraction evolution models both the linear and nonlinear bias
do not differ significantly. But when we compare our FID model to that where all galaxies carry neutral
hydrogen, i.e., $f_B=1$, then the differences are more significant, considering a realistic
blue galaxy fraction lowers the $\HI$ bias, since the neutral hydrogen avoids accumulating in the bigger halos
and concentrates more in mid-size and small mass halos.

The weak dependence of the $\HI$ bias on the model adopted for the
neutral hydrogen evolution means that the intensity mapping
observations will be able to provide only a limited amount of
information on the global evolution of the neutral hydrogen abundance
in the universe. On the other hand, from a cosmologist's point of
view, this is good news: a detailed knowledge of the $\HI$ evolution
may not be needed for studying the dark energy evolution in the radio
intensity mapping experiments. Note that even though the differences
in $b_2$ between the PME and PNE models at high redshift are larger
than those in $b_1$, measurements on higher-order correlations have larger
uncertainties which would make the $b_2$ estimations less reliable. Nevertheless,
if good signal-to-noise measurements can be done, this parameter can
play a relevant role in determining the right evolution model.


\subsection{Linear Bias calculated from simulations}
\label{section:simuls}

In order to have a consistency check of our analytical  calculations,
we turn to estimate the linear $\HI$ bias from the Millennium Simulation
\citep{springel_etal:05}. The N-body simulation was carried out for 2160$^3$ particles with mass
$m_p=8.6\times10^8$ $\hMsun$ in a
box with  $L_{\rm box}=500$ $h^{-1}$Mpc on the side, with the same
cosmological parameters we use for our analytical estimates.

Following Equation (\ref{eq:bias-pcf}), we can estimate
the linear bias in the simulation by
\begin{equation}
 b^{\rm{MS}}_{\HI,1}(k)=\sqrt{\frac{P_\HI(k)}{P_{\rm dm}(k)}},\label{eq:P_ratio}
\end{equation}
where $P_{\rm dm}(k)$ is the real-space dark matter power spectrum and $P_\HI(k)$ is the real-space HI power spectrum.
In addition to validate our analytical method, this measurement should allow us to estimate the
scales below which nonlinear effects become relevant.


We use the
$\Mhi-\Mtot$ relation obtained in \S\ref{section:h21mf} to assign the neutral 
hydrogen mass to dark matter halos and calculate the $\HI$ power spectrum 
$P_{\HI}$. Note that due to the mass resolution of the simulation, our dark matter
halo catalog does not resolve well the halos below a certain mass threshold, and those halos
may contribute substantially to the
value of the $\HI$ bias. Therefore, consistency requires the bias measured in the Millennium
Simulation to be compared to the value obtained
analytically through Equation (\ref{eq:bhi}) assuming the \textit{same} 
mass threshold $M_{\rm min}$. 
For this reason, we compare analytical estimates and measured values of the bias obtained by 
assuming $M_{\rm min}\sim10^{11}h^{-1}M_\odot$, which should correspond to halos with 
$N_p \sim 100$ particles. 
Another aspect when we are ``painting'' $\HI$ in the simulation is its distribution 
inside the halos. In our case, we place all
hydrogen at the center of the halo, i.e we are not assuming diffuse $\HI$ emission inside.
While this might not model accurately the small-scale clustering of the $\HI$, 
what matters in the large-scale clustering are the details of  the  $\Mhi-\Mtot$ relation; the spatial profile of the 
neutral hydrogen within a halo will have a very limited effect on the measured bias and its scale dependence on 
large scales we present in this paper \citep{cooray_sheth:02,Schulz:06}.

\begin{figure}
\plotone{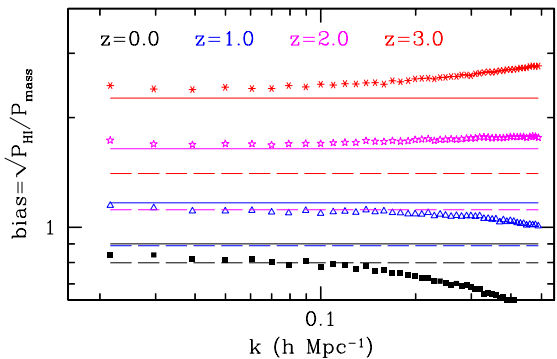}
\caption{\label{fig:biassim}Neutral hydrogen bias factor $b^{\rm{MS}}_{\HI,1}(k)$ at different redshifts
 from the Millennium Simulation (symbols) for the model A 
in the PNE scenario, and our analytical
 predictions for the linear bias $b_{1,\HI}$ (horizontal lines). The
 dashed lines show analytical predictions for the whole set of halos
 ($M_{\rm min}=10^4$ $\hMsun$), while solid lines show the analytical calculations for
 the set of halos resolved in the Millennium Simulation
 ($M_{\rm min}=10^{11}h^{-1}M_\odot$).}
\end{figure}
Figure \ref{fig:biassim} shows, in symbols, the measured bias $b^{\rm{MS}}_{\HI,1}(k)$ for 
a series of redshifts from $z=0$ to $z=3$ for the model A 
in the PNE scenario. The solid lines represent $b_{\HI,1}$ from our analytical
modeling, with a minimum
mass cut $M_{\rm min}=10^{11}$ $\hMsun$; the dashed lines are the results of $b_{\HI,1}$ with
$M_{\rm min}$=$10^4$ $\hMsun$.

The bias measured from the simulation using 
Equation (\ref{eq:P_ratio}) should agree with $b_{\HI,1}$ obtained from Equation (\ref{eq:bhi}) on large scales;
on smaller scales, a contribution from the clustering within a halo results in a scale dependence in bias \citep{Seljak:01} 
which we refer to as a nonlinear bias effect. From the figure, we indeed observe a strong redshift evolution of large-scale bias in agreement with the behavior found in our analytical calculations;
the agreement is of the order of 5\%. The small differences are due mainly to the fact that our analytical halo mass function, though is a good approximation, indeed has small differences 
to the one measured in simulations. The figure also shows that the linear bias assumptions break down for $k \gtrsim
0.15\,h$Mpc$^{-1}$, but the effective nonlinear scale (where the linear bias deviates strongly from a constant value)
appears to evolve little with redshift, likely a consequence of our adopted $M_\HI - M_{tot}$ relationship, given
the fact that in simulations this scale varies greatly with redshift.
We point out that if our assumption of the one-to-one correspondence between the HIPASS 
source and the halo breaks down for large halo masses, it will alter $b^{\rm{MS}}_{\HI,1}(k)$ to some extent. 

As stated earlier, the difference between the two sets of horizontal
lines in Figure \ref{fig:biassim} shows the contribution of
halos with $\Mtot < M_{\rm min}=10^{11}h^{-1}M_\odot$. This contribution is substantial. It is important to note that in the 
case of the 21 cm intensity mapping, 
we suffer no lower detection limit in neutral hydrogen mass and therefore halo mass. 
In that case, the bias derived with $M_{\rm min}=0$ is the appropriate one.
As a much higher
mass resolution would be required to reduce $M_{\rm min}$ much below $10^{11}h^{-1}M_\odot$, the Millennium Simulation --or another
simulation with similar resolution-- cannot be used to
provide an accurate estimate for the $\HI$ bias for this kind of survey, 
but only as a useful tool to validate the semi-analytical estimates obtained in the previous section.

\subsection{Impact of bias evolution on the power spectrum}

Having shown how the $\HI$ bias evolves with redshift, we turn to
assess to what extent this may or may not have an effect on the
measurement of the power spectrum. At tree-level in perturbation
theory, the density fluctuations in the neutral hydrogen $\delta_\HI$ 
and in the dark matter $\delta_{dm}\equiv\delta$ are related through
\begin{eqnarray}
\delta_\HI(\vec{x},\chi)&=&b(\chi)\delta(\vec{x},\chi)
=b(\chi)D_+(\chi)\delta(\vec{x},0)\nonumber\\
&\equiv&E(\chi)\delta(\vec{x},0),
\end{eqnarray}
where $D_+(\chi)$ is the linear growth factor, and we exchanged
redshift $z$ for comoving distance $\chi$. When observations are
performed, only the product $E(\chi)$ can be measured, as the
redshift evolution of the bias contributes a component that combines
with the growth of structure. To understand to what
extent this has an influence on the measurement of the power
spectrum, we start by Taylor expanding $E(\chi)=\sum E_0^{(n)}\chi^n/n!$, where all derivatives $E_0^{(n)}=d^nE/d\chi^n$ are evaluated at the observer's position. Next, we Fourier transform $\delta_\HI(\vec{x})$ to obtain
\begin{eqnarray}
\delta_\HI(\vec{k},\chi)&=&\int d^3\vec{x}\,e^{-i\vec{k}\cdot\vec{x}}\delta_{\HI}(\vec{x},\chi)\nonumber\\
&=&\int d^3\vec{x}\,e^{-i\vec{k}\cdot\vec{x}}\left[\sum_{n=0}^{\infty}\frac{E_0^{(n)}\chi^n}{n!}\right]\delta(\vec{x},0) \nonumber\\
&=&\sum_{n=0}^{\infty} \frac{i^n E_0^{(n)}}{n!} \partial_{k_{\parallel}}^n\delta(\vec{k},0),
\end{eqnarray}
where we used the fact that $\chi^{n}=i^n\,\partial_{k_{\parallel}}^n\,e^{-i\vec{k}\cdot\vec{x}}$.
Next we note that $\delta(\vec{k},0)$ is a Gaussian random variable with dispersion 
$\sigma_{\delta(\vec{k},0)}=\sqrt{P(k,0)}$. We can then rewrite it as 
$\delta(\vec{k},0)=\sqrt{P(k,0)}\lambda_{\vec{k}}$ with $\lambda_{\vec{k}}$ a Gaussian random variable with $\langle\lambda_{\vec{k}}\rangle=0$ and 
$\langle\lambda_{\vec{k}}\lambda^*_{\vec{k}'}\rangle=(2\pi)^3\delta_D^3(\vec{k}-\vec{k}')$. 
Note that $\vec{k}$ for $\lambda_{\vec{k}}$ is nothing but a label and that all the dependence 
of $\delta(\vec{k},0)$ on the wavevector is really encoded in the prefactor. 
Defining the $\HI$ power spectrum by 
$\langle\delta_\HI(\vec{k},\chi)\delta^*_\HI(\vec{k}',\chi)\rangle\equiv P_\HI(k,\chi)\,(2\pi)^3\,\delta_D^3(\vec{k}-\vec{k}')$ 
we can then express $P_\HI$ as
\begin{equation}
P_{\HI}(k,\chi)\approx\left[L(k)+\mu^2 H(k)\right]P(k,0),
\end{equation} 
where $\mu=\cos(\theta)$ is the cosine of the angle the wavevector $\vec{k}$ makes with the line of sight 
and $L(k)$ and $H(k)$ 
are the first two terms that encode the $\chi$ dependence of $\delta_{\HI}$.
We calculate the first few terms of $L$ and $H$ using the fact that $k_{\parallel}=k\cos(\theta)=k\mu$ 
and $\vec{k}_{\perp}=k\sin(\theta)=k\sqrt{1-\mu^2}$. They turn out to be 
\begin{eqnarray}
L&\approx&E_0^2-E_0 E_0''\frac{P'}{2kP},\label{eq:L}\\
H&\approx&E_0E_0''\left(\frac{P'^2}{4P^2}-\frac{P''}{2P}+\frac{P'}{2kP}
\right)+E_0'^2\frac{P'^2}{4P^2},\label{eq:H}
\end{eqnarray}
where $P'$ and $P''$ denote first and second derivatives of $P$ with respect to $k$, while $E_0'$ and $E_0''$ denote derivatives of $E(\chi)$ with respect to the comoving distance evaluated at the observer position. In general, 
the comoving distance dependence generates in the power spectrum a directionality dependence that 
is reminiscent of redshift space distortions. Despite that the nature of the effect considered here is completely different from the one of redshift space distortions (which is due to peculiar velocities), the two effects may be in practice difficult to separate from one another, unless we have a sufficiently small sample variance on the clustering measurement.
However, we point out that the $n$-th derivative with respect to the comoving distance scales 
as $H_0^{n}$ and therefore $E_0'\sim10^{-4}$ and $E_0''\sim 10^{-8}$. 
As such, all but the first term appearing in Equations (\ref{eq:L}) and (\ref{eq:H}) are of order $10^{-8}$ and only the $E_0^2$ term appearing in $L$ will give a non-negligible contribution. Also, it is possible to note that the above treatment is valid even in the case of a perfect unbiased dark matter tracer.
\begin{figure}
\plotone{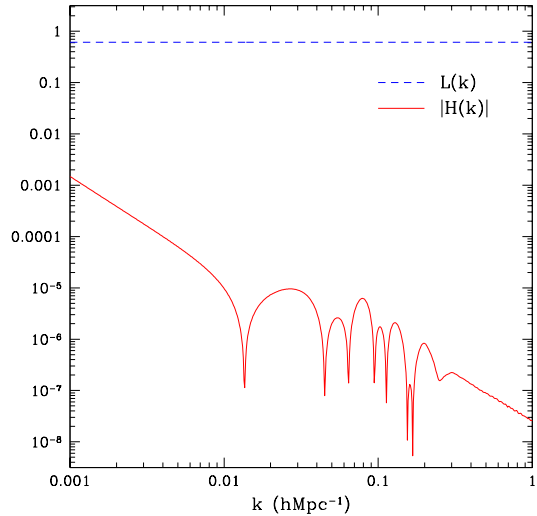}
\caption{\label{fig:PS_corr} First two correction coefficients $L(k)$ and $|H(k)|$.}
\end{figure}

In Figure \ref{fig:PS_corr} we show the functions $L(k)$ and $|H(k)|$.
Consistent with the above estimates, $H$ turns out to be extremely small. On the other hand, $L$ is 
dominated by the term $E_0^2$ and therefore does not show any appreciable $k$ dependence, 
except for extremely large scales $k\sim10^{-3}$, where it deviates from $E_0^2$ by about 0.3\%. 
These two facts together allow us  to conclude that the measurement of the 21 cm power 
spectrum through intensity mapping
should not be sensibly affected by the redshift dependence of the bias.

\section{Summary and Conclusions}
\label{section:conclusions}

In this paper we have estimated analytically and in the Millennium
Simulation the large-scale 
bias of neutral hydrogen from a non-parametric model
which establish a one-to-one relation between the
total mass  and the $\HI$ mass of galactic halos. Although
only an approximation, this relation is expected to hold
for a large range of masses.

The bias parameters have been
calculated using a model which uses elements of the HOD formalism, in a way that can be considered
complementary to the galaxy-HOD model of $\HI$ galaxies carried out by
\cite{wyithe_etal:09}; we do not deal in our case with the individual
galaxies inside dark matter halos but with the total neutral hydrogen
inside them. In terms of large-scale clustering, these approaches are 
completely equivalent.

Using this  approach together with different models for the
redshift evolution of the $\HI$ content in the universe and the
relation between neutral hydrogen and total matter contents of
galactic halos, we analyze the evolution of bias with redshift. We
find that both linear and nonlinear bias increase significantly with
redshift. This contrasts with the finding by \citet{Wyithe:2007rq},
based on an extrapolation to lower redshifts from a reionization model
of \citet{Wyithe:2007gz}. Since \citet{Wyithe:2007rq} do not account for
clustering of galactic halos that host all neutral hydrogen after
reionization, they underestimate the $\HI$ bias at $z \lesssim 4$.
On the other hand, our results are in  agreement with the high value of the 
linear large-scale bias at redshift $z\sim 3$ measured in the model by \cite{bagla_khandai:09}
where they populate $\HI$ in high-resolution $N-$body simulations.

We also find that, given the limitations of our model at 
high masses, the estimation of the bias does not change significantly
when the very high mass halos are neglected.

Our main conclusion is that, despite a wide range of possible models for the
evolution of $\HI$ mass function that we consider, in all models the
bias evolution is similar --the neutral hydrogen distribution is
mildly antibiased at $z=0$ but becomes strongly biased ($b_\HI \sim
2$) by $z \sim 4$. This result is encouraging for the planned radio intensity
mapping experiments, since large bias implies a stronger 21 cm signal.
Nevertheless, this strong redshift evolution does not significantly
compromise the measurements of the neutral hydrogen power spectrum
along the radial direction.

\acknowledgements

We thank Gerard Lemson for his support and patience in providing the
Millennium Simulation halo catalogs.  We also thank Andrey Kravtsov for
his comments and suggestions.

This work was supported in part by the Kavli Institute for Cosmological Physics 
at the University of Chicago through grants NSF PHY-0114422 and NSF PHY-0551142 
and an endowment from the Kavli Foundation and its founder Fred Kavli.
AV and HS are supported by the DOE at Fermilab. The Millennium Simulation databases used in this paper and the web application providing online access to them were constructed as part of the activities of the German Astrophysical Virtual Observatory.

\bibliography{biash21,ng-bibs/igm,ng-bibs/qlf,ng-bibs/self,ng-bibs/21cm,ng-bibs/misc}

\end{document}